\documentclass[]{spie}  

\usepackage{amsmath,amsfonts,amssymb}
\usepackage{graphicx}
\usepackage[colorlinks=true, allcolors=blue]{hyperref}

\title{Cold Stop and Lyot Stop Designs for a New Infrared Exoplanet Imager at Keck Observatory}

\author[a]{Jialin Li}
\author[b]{Andrew Skemer}
\affil[a]{Department of Physics, University of California Santa Cruz, 1156 High St, Santa Cruz, CA 95064, USA}
\affil[b]{Department of Astronomy and Astrophysics, University of California Santa Cruz, 1156 High St, Santa Cruz, CA 95064, USA}

\authorinfo{Further author information: 
\\Jialin Li: E-mail: jli428@ucsc.edu
\\Andrew Skemer : E-mail: askemer@ucsc.edu}

\begin{document} 
\maketitle

\begin{abstract}
Santa Cruz Array of Lenslets for Exoplanet Spectroscopy (SCALES) is an instrument being designed for direct imaging of exoplanets in the infrared with the Adaptive Optics System of the W.M. Keck Observatory. The performance of SCALES will be largely affected by thermal transmission and emission from various sources, including the adaptive optics and instrument structures. The placement of a cold stop and a Lyot stop can preserve maximal and stable throughput while limiting the emission of instrument structures such as primary mirror segment gaps, secondary structures, and spider arms. Here we propose and compare three cold stops, a circular inner mask paired with circular, hexagonal, and serrated outer masks, as well as one Lyot stop design. Taking into account the pupil nutation and mirror emissivity, we model the throughput and the background emission for all designs to optimize the dimensions of the cold stop and the Lyot stop.
\end{abstract}


\section{INTRODUCTION}
\label{sec:intro}
For infrared exoplanet imaging instrumentation, cold stops and Lyot stops are essential for suppressing unwanted thermal emission from the telescope structure, and the host star. The cold stop is designed to obscure all unwanted radiation while preserving the maximal amount of signal, and the Lyot stop is designed such that there is a stable amount of signal input. Ideally, the masks will match the dimension and shape of the telescope pupil, blocking all thermal radiation and scattered light not originated from the primary mirrors, including the secondary support structures and the segment gaps. However, with physical limitations, misalignments, and pupil nutation, the ideal masks are unlikely the most suitable. 

In this proceeding, three cold stop designs and one Lyot stop design are proposed with pupil nutation taken into account via modeling the throughput and the background emission for the Santa Cruz Array of Lenslets for Exoplanet Spectroscopy (SCALES). SCALES is a thermal-infrared diffraction limited imager and high-contrast lenslet integral field spectrograph (IFS) being designed for direct imaging and characterizing exoplanets at the W.M. Keck Observatory \cite{2020SPIE11447E..64S, 2020SPIE11447E..4ZB}. The pupil of the Keck II telescope is described in Section~\ref{sec:keck}, the cold stop designs are discussed in Section~\ref{sec:coldstop}, the Lyot stop design is presented in Section~\ref{sec:lyotstop}, and the effects of degree of pupil nutation and telescope emissivity is assessed in Section \ref{sec:conclusion}.

\section{PUPIL OF THE KECK II TELESCOPE} 
\label{sec:keck}
\begin{figure} [htbp] 
\begin{center}
\includegraphics[height=12cm]{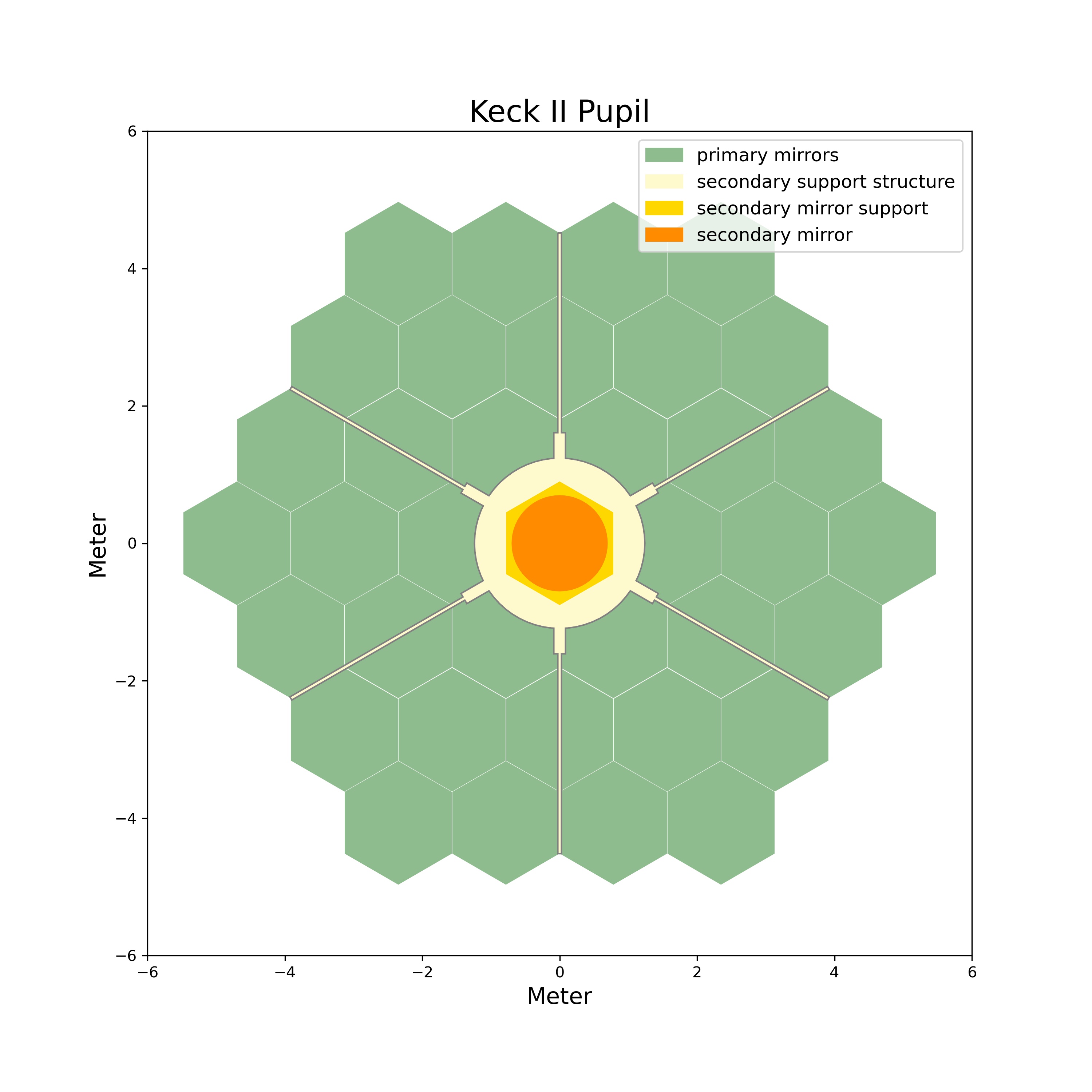}
\end{center}
\caption {Diagram of the Keck II pupil. The secondary support structure includes all components in the plane of the secondary mirror. The secondary mirror and mirror support are shown explicitly for clarity.\label{fig:pupil}}
\end{figure}

The primary mirror of the Keck II telescope consists of 36 regular hexagonal segments of side length 0.9 m with a segment gap of 0.003 m between each segment. A diagram of the Keck pupil is shown in Figure~\ref{fig:pupil}. An additional non-reflected gap width of 0.002 m is located on each side of the segment gaps \cite{LeMignant96,Nelson85}.

When viewing an image of the telescope pupil, secondary structures will also be seen, including six secondary supports, the secondary mirror, and the secondary obscuration. The secondary mirror has a diameter of 1.4 m\cite{Nelson86}. The diameter of the secondary obscuration was measured to be 2.6 m\cite{Nelson85} and 2.48 m via images taken by the NIRC2 camera \cite{2016SPIE.9908E..35A}. The spider arms supporting the secondary structure have a uniform width of 0.025 m. The six secondary support node connects the secondary obscuration and the spider arms with a larger wide than the spider arms. From the same images taken by the camera of NIRC2, nodes of the spider arms are measured to have a length of 0.3692 m and a width of 0.08m \cite{arriaga-pc,lyke-pc}. 

Our model of the Keck II pupil does not take into account the additional non-reflective gap and considers it to be a functional section of each mirror segment as they only account for 0.7\% of the total primary area\cite{Nelson85}. The emissivity of the warm optics (primary/secondary/tertiary mirrors + adaptive optics system) has been measured to be 30\%\cite{hinz-pc}, and it is the value used for the following calculations. The diameter of the secondary obscuration is assumed to be 2.48 m as it is a value measured more recently. Additionally, its value is evaluated from the same measurement and method as the dimensions of the secondary support nodes. 

\subsection{Pupil Nutation}
Due to the alignment precision of the adaptive optics K-mirror and the Keck telescope beam, the Keck II pupil nutation has been measured to be approximately 1\% of the diameter of the primary mirror\cite{2016SPIE.9909E..22F}. Neglecting the effect of nutation will result in decreased overlap between the pupil and the cold-stop, and variation in the point spread function (PSF). Although the nutation was measured, without regular maintenance, nutation can get worse with time \cite{mawet-pc}. Therefore, when designing the Lyot stop, we adapt a conservative value of 2\%. Additionally, the variation of signal to noise ratio (SNR) from the range of 0\% to 5\% nutation is examined.

\begin{figure} [tp] 
\begin{center}
\includegraphics[height=6.cm]{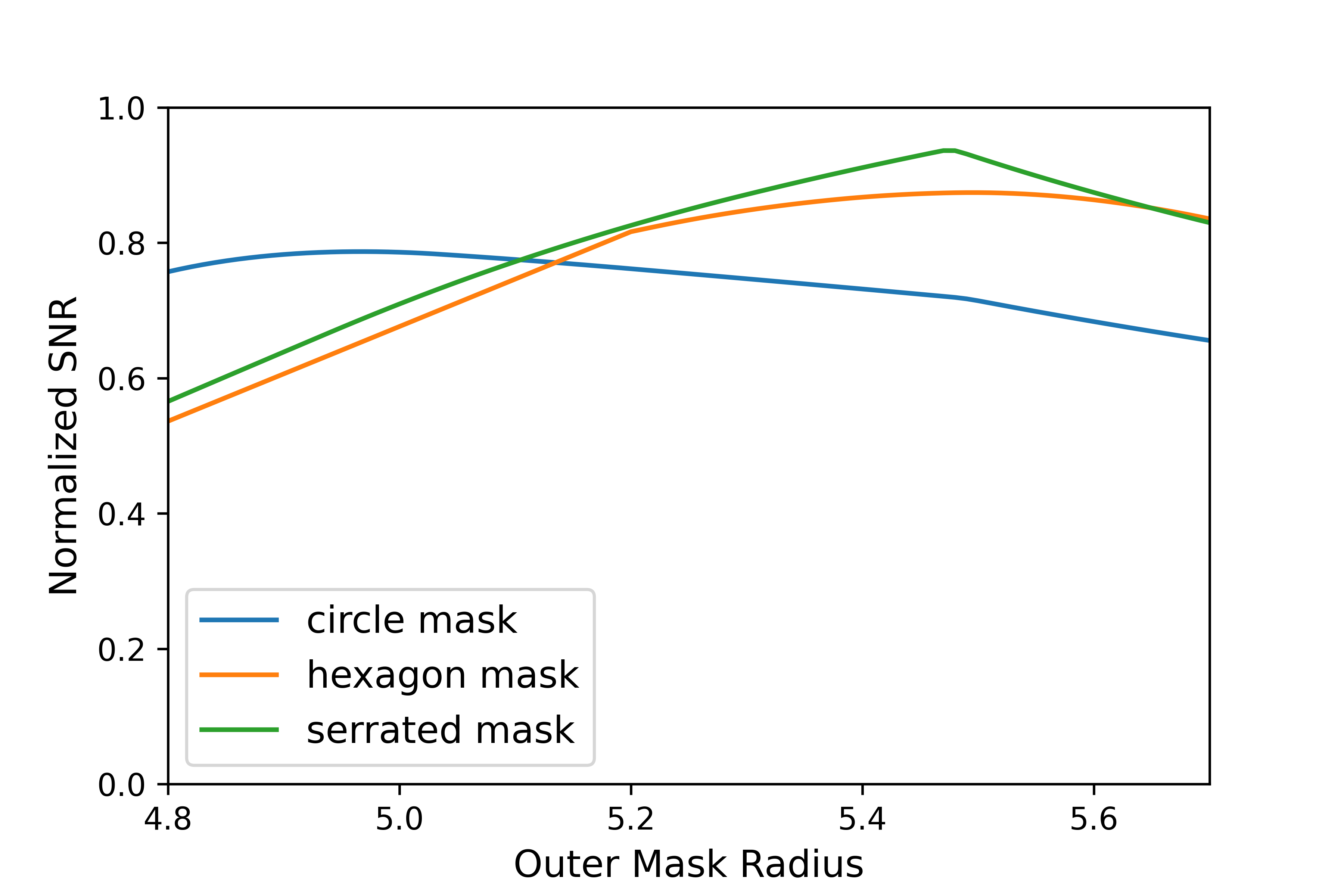}
\end{center}
\caption {This figure shows the normalized SNR for the 3 different outer masks of the cold stop when paired with a circular inner mask. The SNR peaks at 4.97 m, 5.49 m, and 5.47 m respectively for the circular, hexagonal, and serrated hexagonal outer masks.\label{fig:SNR}}
\end{figure} 

\section{COLD STOP DESIGN} \label{sec:coldstop}
Attempting to block all segment gaps between the primary mirrors would only decrease the SNR, as more signal than noise will be blocked due to the physical limitation to the minimum manufacturing size of physical features on the cold stop. Therefore, the background radiation from the segment gaps is disregarded for the optimization.

The cold stop consists of three parts, an inner mask, an outer mask, and spider arms. A circular inner mask with a radius of 1.24 m would be optimal for a still Keck telescope as it matches the shape and size of the secondary obscuration, and it best occludes noises from the secondary structures based on the SNR calculated. 

For the outer mask, a total of three shapes were optimized and compared: circular, hexagonal, and serrated hexagonal. Figure \ref{fig:SNR} shows the normalized SNR for each outer mask. Among the three outer masks, the serrated hexagonal mask with a radius 5.47 m returns the highest SNR as it matches the shape and size of the Keck primary mirrors. Unlike the inner mask and serrated hexagonal mask, we considered oversized and undersized circular and hexagonal masks; the difference being that the oversized masks preserve all light from the primary mirror, while the undersized masks optimize the SNR.

\begin{figure} [tp] 
\begin{center}
\includegraphics[height=6.cm]{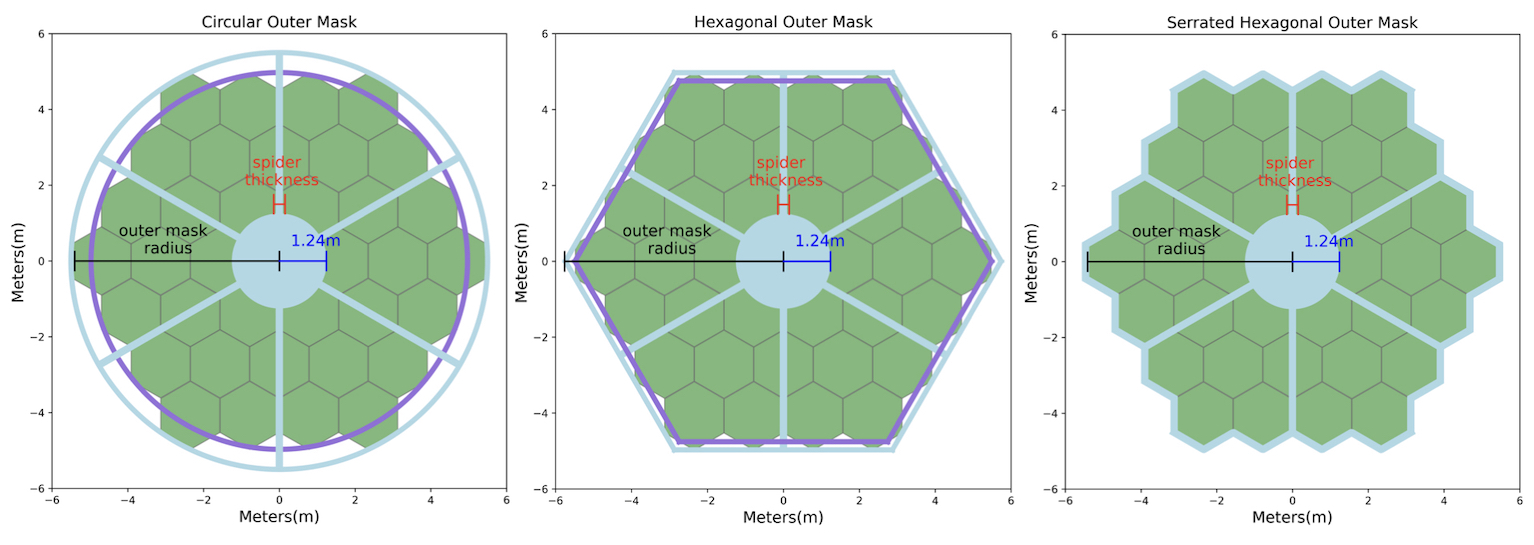}
\end{center}
\caption {Optimal oversized and undersized outer mask radii. The optimal undersized radii are in purple while the optimal oversized are in blue. For the serrated hexagonal outer mask, the optimal radius is the radius that returns the maximal SNR, so only the blue mask outline is present in the figure.\label{fig:dif_shapes}}
\end{figure}

For the circular outer mask, the optimal oversized and undersized radius is 5.5 m and 4.97 m respectively, while for the hexagonal mask, they are 5.74 m and 5.49 m. Figure \ref{fig:dif_shapes} shows the optimal radii for all the different outer masks. Of the various possibilities, the serrated hexagon enables the highest SNR.

For the optimization of the spider arms, the widths are uniform throughout with no nodes present. Since the normalized SNR decreases with increasing width, the spider arm width of 0.026 m, closest width to the secondary supports, returns the peak SNR. However, taken into consideration the secondary support node, a spider arm width of 0.087 m would occlude all noises from the secondary structures. The optimal cold stop design for a still telescope will be a serrated hexagonal outer mask of radius 5.48 m paired with a circular inner mask of radius 1.24 m, along with spider arms of width 0.026 m at the telescope. 

\begin{figure} [tp] 
\begin{center}
\includegraphics[height=5cm]{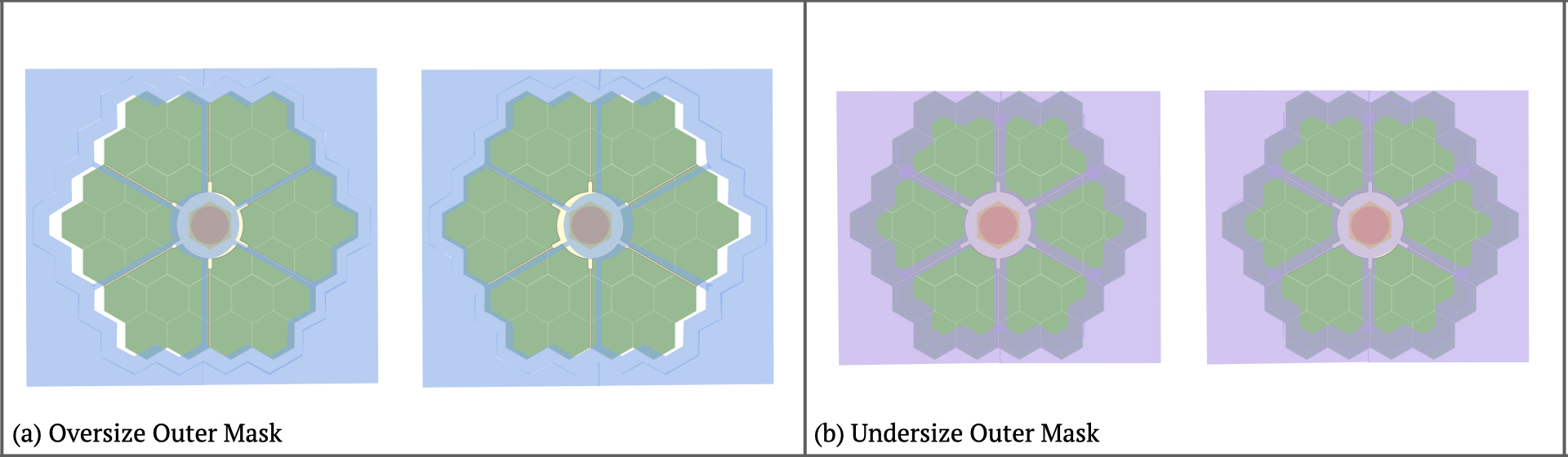}
\end{center}
\caption {Visualization of the throughput of (a)oversized and (b)undersized masks. The blue and purple rectangles represent the masks and the green patches represent the Keck primary mirrors. (a)This oversized design is adopted for the cold stop as the majority of the signal from the primary mirrors is preserved. Noise from the background is shown as the white gap between the mirrors and the mask. (b)This undersized design is adopted for the Lyot stop; Although some parts of the primary mirrors are cut off, throughput remains constant with nutation. \label{fig:under vs over}}
\end{figure}

Due to nutation, the image of the pupil inside the cold stop could shift, causing a section of the image to be cut off with the optimal cold stop design. For SCALES, we plan to adopt an oversized cold stop to preserve as much of the telescope pupil as possible (see Figure \ref{fig:under vs over}a) along with an undersized Lyot stop for a stable coronagraphic PSF (see Section \ref{sec:lyotstop}).

\begin{figure} [tp] 
\begin{center}
\includegraphics[height=6.cm]{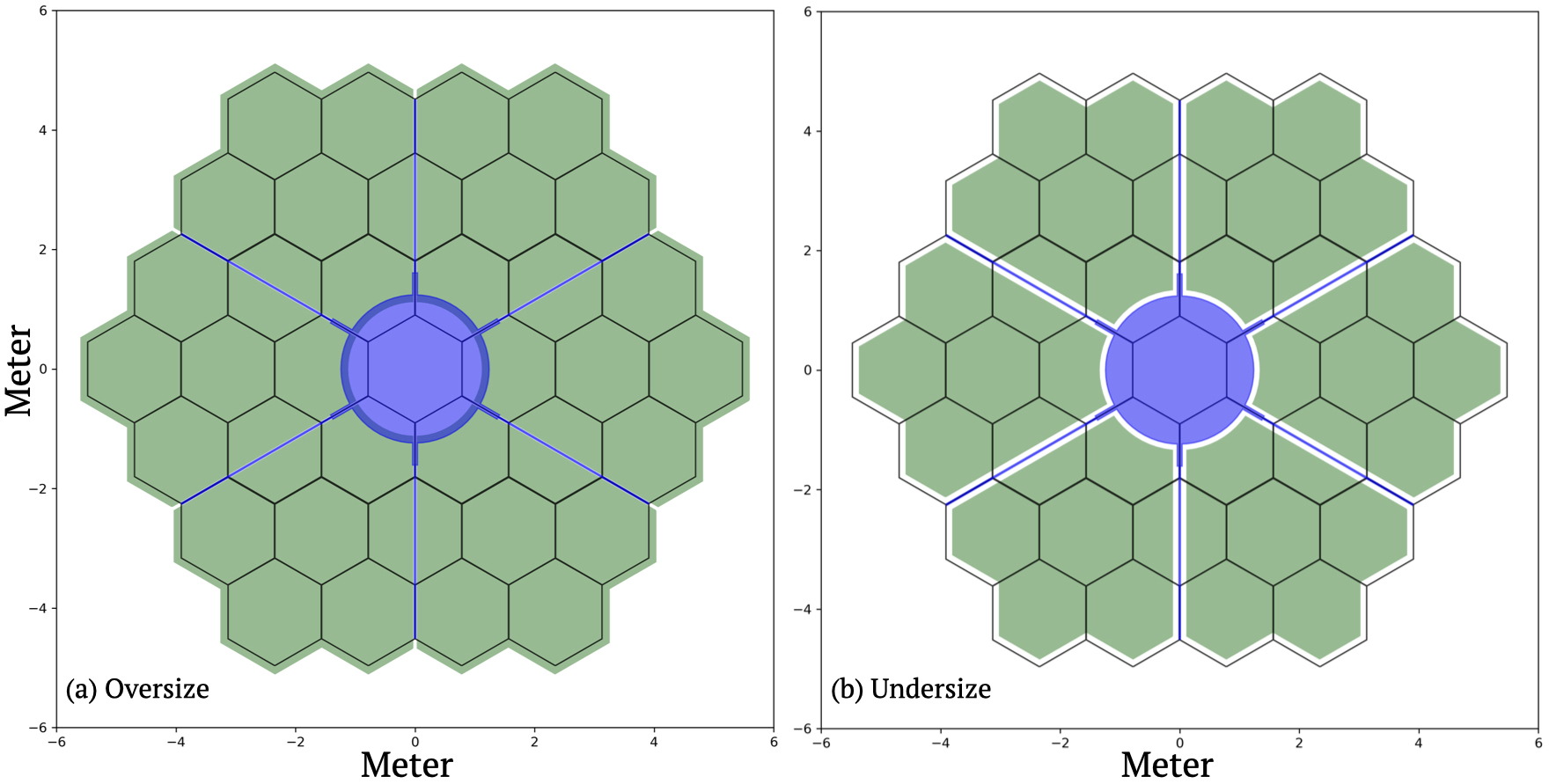}
\end{center}
\caption {The optimal designs of the (a)cold stop and (b)Lyot stop. The secondary structures are shown in blue and the outlines of the primary mirrors are shown in black. The green patches are representations of the carved-out sections in which the light can pass through. (a)The oversized cold stop design consists of a circular inner mask of radius 1.13 m, spider arms of width 0.08 m, and a serrated hexagon outer mask of 5.63 m. The overlapping region of the green patches and the secondary structures located around the secondary obscuration will allow light to pass through when the pupil is shifted. (b)The undersized Lyot stop design consists of a circular inner mask of radius 1.35 m, spider arms of width 0.245 m, and a 5.32 m. \label{fig:opt_designs}}
\end{figure}

At 2\% nutation, the length of the movement is 0.1 m, subsuming both the spider arm and the spider arm nodes. At this nutation, the spider arms and nodes can be treated as one and the optimal width of the undersized spider arms is 0 m, which is not feasible to manufacture for the steel blockers that we plan to use with SCALES. To achieve the goal of preserving the maximum amount of signal while blocking unwanted emission, the optimal spider arm width is assumed to be the same width of the secondary support nodes, 0.08 m, and remains constant with nutation. Thus, the optimal cold stop at 2\% nutation consists of a serrated hexagonal outer mask of radius 5.63 m, a circular inner mask of radius 1.13 m, and spider arms of width 0.08 m. Figure \ref{fig:opt_designs}a shows the design of such a cold stop. With such a cold stop design, more background noise will be detected with increasing degree of nutation, resulting in the decrease of SNR; the SNR of the optimal cold stop at 2\% nutation is 89\% of that for a still telescope as shown in Figure \ref{fig:SNR change}a. 

\section{LYOT STOP DESIGN} \label{sec:lyotstop}
Similar to the cold stop, the Lyot stop also consists of three parts, and the optimal shapes for the inner and outer masks remain to be circle and serrated hexagon respectively. Unlike the cold stop, the design of the Lyot stop aims to stabilize the input with an undersized outer mask and oversized inner obscuration (inner mask and spider arms), such that the pupil does not change with nutation (see Figure \ref{fig:opt_designs}b). Consequently, the SNR of the optimal cold stop design will not be the largest as parts of the image will not pass through the Lyot stop as shown in Figure \ref{fig:under vs over}b.

\begin{figure} [tp] 
\begin{center}
\includegraphics[height=5cm]{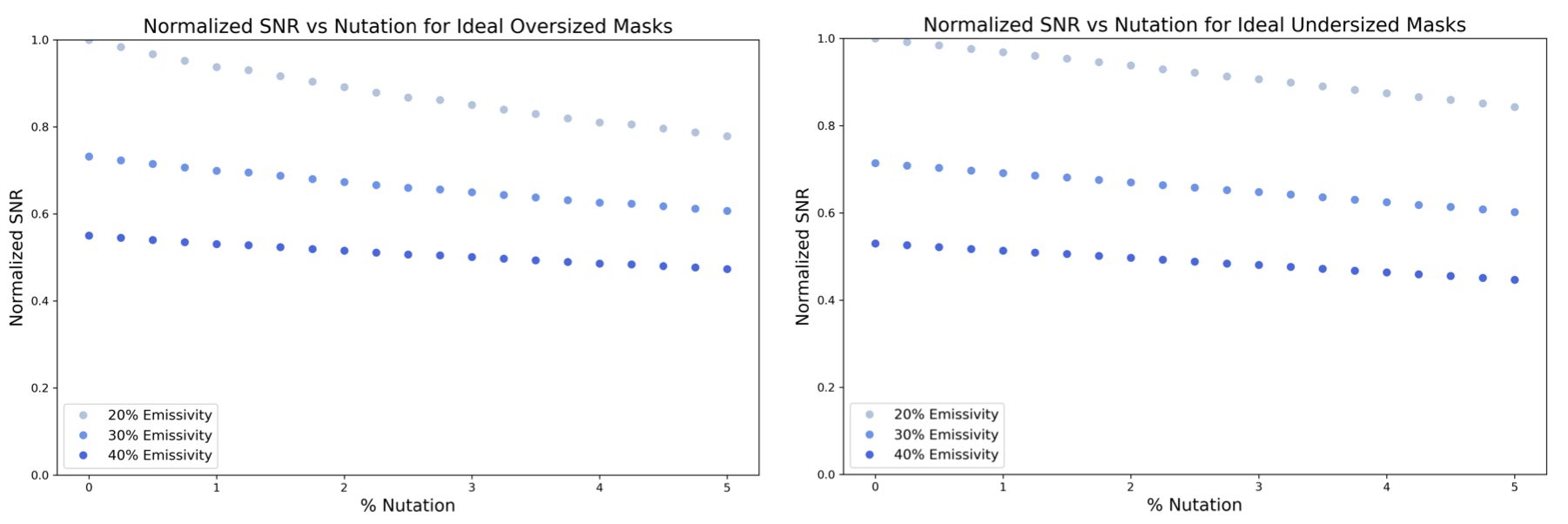}
\end{center}
\caption {Change in SNR vs nutation for the optimal designs with (a)oversized mask and (b)undersized mask for 3 different emissivities. As the nutation increases, in percentage of the primary mirror diameter, the normalized SNR decreases for both. At 2\% pupil nutation, the peak SNR  drops (a)11\% and (b)8\% when compared values returned in the still telescope case. With the cold stop, SNR of the 30\% and 40\% emissivity cases, are respectively 73\% and 55\% of the 20\% emissivity case; With the Lyot stop, SNR of the 30\% and 40\% emissivity cases, are respectively 71\% and 53\% of the 20\% emissivity case. \label{fig:SNR change}}
\end{figure}

Again, assuming that the pupil nutates at 2\% of the diameter of the primary mirrors, the optimal values for the serrated hexagon outer radius, circular inner mask radius, spider arm width are 5.32 m, 1.35 m, and 0.245 m. A figure of the optimal Lyot stop is shown in Figure \ref{fig:opt_designs}b. The inner mask radius and spider arm widths are unchanged because they serve to block the radiation from the secondary structures and the telescope spider arms. The SNR at 2\% nutation decreased to approximately 8\% of the SNR with an optimal Lyot stop design for a still telescope, which is the same design as the optimal cold stop, as shown in Figure \ref{fig:SNR change}b.

\section{CONCLUSION} 
\label{sec:conclusion}
In this paper, a cold stop and a Lyot stop for SCALES have been designed via the calculation and analysis of the maximal SNR and stable PSF in relation to the mask dimensions for optimal performance. There are a few factors that can improve the performance of the instrument and the telescope. 

We have assessed the effect of the primary mirrors emissivity, which is assumed to be 30\% for previous calculations, on SNR of the optimal cold stop and Lyot stop designs. The normalized SNR values of optimal mask designs with three different emissivities are shown in Figure \ref{fig:SNR change}. As nutation of the pupil increases, the SNR of optimal mask with 30\% emissivity remains to be about 55\% of the SNR value of the same oversize mask with 20\% emissivity and 53\% of the same undersized mask. Similarly, a mask with 40\% emissivity returns about 73\% of the SNR value of the same oversized mask with 20\% emissivity and 71\% of the same undersized mask. The decrease in the emissivity can be achieved (i.e. cleaning the primary mirrors), and the performance of the telescope can be greatly enhanced. Additionally, alignment of the K-mirrors can eliminate the 2\% pupil nutation, which will result in improvement of SNR by approximately 8-11\% and therefore a much more stable PSF. Manufacturing tolerances would be taken into account during the fabrication of the cold stop and Lyot stop, as they influence the robustness and performance of the actual masks. The optimal Lyot stop described in the previous section is designed without taking into account diffraction. The effects of diffracted light can also be a limiting factor to the performance of the mask and will require further assessment.

\acknowledgments 
This project was made possible by a Lloyd B. Robinson Undergraduate Research in Instrumentation Award from UC Santa Cruz.  We gratefully acknowledge the support of the Heising-Simons Foundation through grant \#2019-1697.  We thank Pauline Arriaga for useful discussions.

\bibliography{main} 
\bibliographystyle{spiebib} 

\end{document}